\documentclass[aps,pra,reprint,amsmath,amsfonts,amssymb,superscriptaddress]{revtex4-1}
\usepackage{graphicx,float,calc}
\usepackage{color,bm}
\usepackage{ulem}
\usepackage{braket}
\usepackage[colorlinks,urlcolor=blue,citecolor=blue,linkcolor=blue]{hyperref}

\makeatletter

\newcommand{\Rmnum}[1]{\expandafter\@slowromancap\romannumeral #1@}
\makeatother

\begin{document}
\title{Extracting non-Abelian quantum metric tensor and its related Chern numbers}

\author{Hai-Tao Ding}
\affiliation{National Laboratory of Solid State Microstructures and School of Physics, Nanjing University, Nanjing 210093, China}
\affiliation{Collaborative Innovation Center of Advanced Microstructures, Nanjing 210093, China}

\author{Yan-Qing Zhu}
\affiliation{Guangdong Provincial Key Laboratory of Quantum Engineering and Quantum Materials, School of Physics and Telecommunication Engineering, South China Normal University, Guangzhou 510006, China}
\affiliation{Guangdong-Hong Kong Joint Laboratory of Quantum Matter, Frontier Research Institute for Physics, South China Normal University, Guangzhou 510006, China}
\author{Peng He}
\affiliation{National Laboratory of Solid State Microstructures and School of Physics, Nanjing University, Nanjing 210093, China}
\affiliation{Collaborative Innovation Center of Advanced Microstructures, Nanjing 210093, China}

\author{Yu-Guo Liu}
\affiliation{National Laboratory of Solid State Microstructures and School of Physics, Nanjing University, Nanjing 210093, China}
\affiliation{Collaborative Innovation Center of Advanced Microstructures, Nanjing 210093, China}

\author{Jian-Te Wang}
\affiliation{National Laboratory of Solid State Microstructures and School of Physics, Nanjing University, Nanjing 210093, China}
\affiliation{Collaborative Innovation Center of Advanced Microstructures, Nanjing 210093, China}

\author{Dan-Wei Zhang}
\email{danweizhang@m.scnu.edu.cn}
\affiliation{Guangdong Provincial Key Laboratory of Quantum Engineering and Quantum Materials, School of Physics and Telecommunication Engineering, South China Normal University, Guangzhou 510006, China}
\affiliation{Guangdong-Hong Kong Joint Laboratory of Quantum Matter, Frontier Research Institute for Physics, South China Normal University, Guangzhou 510006, China}

\author{Shi-Liang Zhu}
\email{slzhu@scnu.edu.cn}
\affiliation{Guangdong Provincial Key Laboratory of Quantum Engineering and Quantum Materials, School of Physics and Telecommunication Engineering, South China Normal University, Guangzhou 510006, China}
\affiliation{Guangdong-Hong Kong Joint Laboratory of Quantum Matter, Frontier Research Institute for Physics, South China Normal University, Guangzhou 510006, China}

\begin{abstract}
The complete geometry of quantum states in parameter space is characterized by the quantum geometric tensor, which contains the quantum metric and Berry curvature as the real and imaginary parts, respectively. When the quantum states are degenerate, the quantum metric and Berry curvature take non-Abelian forms. The non-Abelian (Abelian) Berry curvature and Abelian quantum metric have been experimentally measured. However, an experimentally feasible scheme to extract all the components of the non-Abelian quantum metric tensor is still lacking. Here we propose a generic protocol to directly extract the non-Abelian quantum metric tensor in arbitrary degenerate quantum states in any dimensional parameter space based on measuring the transition probabilities after parameter quenches. Furthermore, we show that the non-Abelian quantum metric can be measured to obtain the real Chern number of a generalized Dirac monopole and the second Chern number of a Yang monopole, which can be simulated in three and five-dimensional parameter space of artificial quantum systems, respectively. We also demonstrate the feasibility of our quench scheme for these two applications with numerical simulations.
\end{abstract}

\date{\today}
\maketitle

\section{Introduction}

Geometry bears numerous fascinating phenomena in many branches of modern physics. In quantum and condensed-matter physics, the complete geometric properties of quantum states in Hilbert space (which can be real, momentum, or parameter space) are characterized by the quantum geometric tensor (QGT) \cite{RJSlager201919,Michael2017,Mayuquan2010,Rezakhani2010,SLZhu2008,DGonzalez2020,GPalumbo20214,AWZhang2021,BMera2021,Resta2011,Berry1989,Nakahara2003,Provost1980,GRigolin2008,GRigolin2014,Grandi2011,Shankar2017,Polkovnikov2013,Rattacaso2020,Zanardi20077,Palumbo20182,OBleu2018,DWZhang2018,DGonzalez2020,GPalumbo20214}. The QGT is generally complex, with real and imaginary parts. For nondegenerate quantum states, the imaginary part of the QGT corresponds to the famous $U(1)$ Abelian Berry curvature \cite{MBerry1989,Berry1984,DXiao2010}, which has been widely explored in theory and experiments. Recently, the Abelian Berry curvature has been directly measured in cold atom systems \cite{ColdAtom1,ColdAtom2,ColdAtom3}, photonic lattices \cite{Wimmer2017}, and superconducting quantum circuits \cite{Schroer2014,Roushan2014,Tan2018,Tan2019}. The Berry curvature has deep connections to the geometry and topology of Bloch bands. For example, with regard to the Weyl semimetals \cite{Armitage2018}, the topological charge of a Weyl point in momentum space is the first Chern number as the integral of the Berry curvature over a $S^2$ sphere enclosing it.

In non-degenerate case, the real part of the QGT defines the Abelian quantum metric, which measures the distance between two neighbor quantum states in the $U(1)$ vector bundle. The quantum metric also plays an important role in various areas, such as the quantum transport \cite{Neupert2013,Kolodrubetz2013,Kolodrubetz2017,Kolodrubetz2017,Albert2016,OzawaT2018,Bleu2018,Lapa2019,YGao2019}, the quantum information theory \cite{Campos2007,Zanardi2007,You2007,Mahapatra2012,Abasto2008,Albuquerque2020}, and the topological states of matter \cite{GPalumbo2018,Legner2013,RRoy2014,Claassen2015,Bauer2016,Kuzmak2018,Palumbo2018}. Several schemes have been proposed to extract the Abelian quantum metric \cite{Kolodrubetz2013,Ozawa2018,DingHT2020,TOzawa2019,LKLim2015,OBleu2018,Neupert2013,RLKlees2020}. Remarkably, the Abelian quantum metric has recently been measured in different engineered quantum systems, which include the superconducting quantum circuits \cite{TanXS2019,TanXS2021}, the nitrogen-vacancy center in diamond \cite{MYu2020,MChen}, the planar microcavity \cite{Gianfrate2020}, and cold atoms in optical lattices \cite{Asteria2019}.

For degenerate quantum states, the complete geometry is characterized by the non-Abelian QGT \cite{Mayuquan2010,Rezakhani2010,DGonzalez2020,GPalumbo20214,AWZhang2021,BMera2021}. When the degenerate states are $N$-fold, the non-Abelian QGTs are naturally defined on the $U(N)$ vector bundle and contain the non-Abelian generalizations of the quantum metric and Berry curvature. Recently, the $U(2)$ non-Abelian Berry curvature has been experimentally measured from the non-adiabatic response effect \cite{Kolodrubetz2016,Abigail2018}, and the related second Chern number of a quantum-simulated non-Abelian Yang monopole \cite{Yang} in the five-dimensional parameter space has been measured. Very recently, the general relations between the (Abelian and non-Abelian) quantum metric and $n$-th Chern numbers in $2n$ spatial dimensions have been revealed \cite{AWZhang2021,BMera2021}. In particular, based on the periodic modulation method \cite{Ozawa2018}, a scheme to extract the second Chern number from measuring the sum of the non-Abelian quantum metric tensor has been proposed \cite{AWZhang2021}. Hitherto, an experimentally feasible scheme to extract all the components of non-Abelian quantum metric tensors for general degenerate quantum states is still lacking.

In this paper, we generalize our quench protocol of quantum-metric measurements in non-degenerate systems in Refs. \cite{TanXS2019,TanXS2021} to more generally degenerate systems. Based on measuring the transition probabilities after parameter quenches, we propose to directly extract all the components of the non-Abelian quantum metric tensor in arbitrary degenerate quantum states in any dimensional parameter space. Furthermore, we show that the non-Abelian quantum metric can be used to obtain the real Chern number of a generalized $Z_2$-type Dirac monopole and the second Chern number of a Yang monopole, which can be simulated in three and five-dimensional parameter space of artificial quantum systems, respectively. We also demonstrate the feasibility of our quench scheme for these two applications with numerical simulations.

The rest of this paper is organized as follows. Section \ref{sec2} is devoted to introducing the Abelian and non-Abelian quantum metric tensor. In Sec. \ref{sec3}, we propose a general quench scheme to extract all the components of the non-Abelian quantum metric tensor. In Sec. \ref{sec4}, we demonstrate our quench scheme to extract the real Chern number and second Chern number from the quantum-metric measurements with numerical simulations. Finally, a brief discussion and a short conclusion are given in Sec. \ref{sec5}.

\section{Quantum metric tensor} \label{sec2}
We begin with a brief review on the Abelian and non-Abelian QGTs for a generic Hamiltonian $H(\boldsymbol\lambda)$ parameterized by $\boldsymbol\lambda=(\lambda_1,\lambda_2,...,\lambda_D)$ in $D$-dimensional parameter space. We suppose that the ground states are $N$-fold degenerate with the eigenstates $\{|\psi_{j}(\boldsymbol\lambda)\rangle\}~(j=1,2,..,N)$, which are separated from $M$ excited eigenstates labeled as $\{|\psi_{N+1}(\boldsymbol\lambda)\rangle,...,|\psi_{N+M}(\boldsymbol\lambda)\rangle\}$ in energy. When $N=1$, the ground state reduces to the non-degenerate one $|\psi_{1}(\boldsymbol\lambda)\rangle$. In this non-degenerate case, the distance between two neighbor states $|\psi_{1}(\boldsymbol\lambda)\rangle$ and $|\psi_{1}(\boldsymbol\lambda+d\boldsymbol\lambda)\rangle$ in the parameter space is given by \cite{Provost1980}
\begin{equation}
\label{ds}
d s^{2}=1-|\langle \psi_{1}(\boldsymbol\lambda) |\psi_{1}(\boldsymbol\lambda+d\boldsymbol\lambda) \rangle|^{2}.
\end{equation}
The adiabatic evolution of the non-degenerate ground state will lead to the Abelian QGT for $|\psi_{1}(\boldsymbol\lambda)\rangle$ \cite{Mayuquan2010,Rezakhani2010}:
\begin{equation}
Q_{\mu \nu}=\langle\partial_{\lambda_{\mu}} \psi_{1}|(1-|\psi_1\rangle\langle \psi_1|)| \partial_{\lambda_{\nu}} \psi_{1}\rangle=g_{\mu \nu}-i F_{\mu \nu}/2,
\end{equation}
which is gauge-invariant and describes the complete geometry of the quantum state manifold. Here all derivatives are taken with respect to the parameters with $\mu,\nu\in\{1,2,...,D\}$. The elements of the Abelian QGT $Q_{\mu \nu}$ take complex values in general. The imaginary and antisymmetric component $F_{\mu \nu}=-2\text{Im} [Q_{\mu\nu}]=-F_{\nu \mu}$ is the Abelian $U(1)$ Berry curvature, while the real and symmetric component $g_{\mu \nu}=\text{Re}[Q_{\mu \nu}]=g_{\nu \mu}$ defines the Abelian quantum metric. The resulting quantum metric tensor $g$ in the Abelian case is the $D\times D$ matrix:
\begin{equation}
g=\left(\begin{array}{lll}
g_{11} &  \cdots  & g_{1D}\\
~~\vdots &  \ddots  & ~~\vdots   \\
g_{D1} &  \cdots  & g_{DD}
\end{array}\right)_{D\times D}.
\end{equation}
The quantum metric characterizes the distance between nearby states in the parameter space and then Eq.~\eqref{ds} can be rewritten as
\begin{equation}
d s^{2}=\sum_{\mu,\nu=1}^{D} g_{\mu \nu} d \lambda_{\mu} d \lambda_{\nu}.
\end{equation}
Note that the distance $d s^{2}$ corresponds to the transition probability of the non-degenerate ground state being
excited to other eigenstates during a sudden quench of the parameter from $\boldsymbol\lambda$ to $\boldsymbol\lambda+d\boldsymbol\lambda$ \cite{Michael2017}. This correspondence provides the way to directly measure the Abelian quantum metric in non-degenerate quantum systems by the sudden quench method as experimentally demonstrated in Refs. \cite{TanXS2019,TanXS2021}.

For the degenerate case of $N\geqslant2$, the QGT becomes the non-Abelian form naturally defined on the $U(N)$ vector bundle of the $N$-fold degenerate quantum states \cite{Mayuquan2010,Rezakhani2010}. Consider a generic state $|\Psi_{0}(\boldsymbol\lambda)\rangle=\sum_{j=1}^{N} c_{j}(\boldsymbol\lambda)|\psi_{j}(\boldsymbol\lambda)\rangle$ under the degenerate ground-state basis $\{|\psi_{j}(\boldsymbol\lambda)\rangle\}~(j=1,2,..,N)$. The distance between $|\Psi_{0}(\boldsymbol\lambda)\rangle$ and its nearby state $|\Psi_{0}(\boldsymbol\lambda+d\boldsymbol\lambda)\rangle$ takes the form
\begin{equation}
\label{nA-distance}
\begin{aligned}
d S^{2}&=1-|\langle \Psi_{0}(\boldsymbol\lambda)|\Psi_{0}(\boldsymbol\lambda+d\boldsymbol\lambda) \rangle|^{2} \\
&=\sum_{\mu,\nu=1}^{D}\langle\partial_{\lambda_{\mu}} \Psi_{0}(\boldsymbol\lambda) | \partial_{\lambda_{\nu}} \Psi_{0} (\boldsymbol\lambda)\rangle d\lambda_{\mu} d\lambda_{\nu}\\
&=\sum_{\mu,\nu=1}^{D}\left[\left(c_{1}^{*},~\cdots,~c_{N}^{*}\right) Q_{\mu \nu}\left(\begin{array}{c}
c_{1} \\
\vdots \\
c_{N}
\end{array}\right)\right] d \lambda_{\mu} d \lambda_{\nu}.
\end{aligned}
\end{equation}
In this case, $Q_{\mu \nu}$ becomes an $N \times N$ matrix with the complex elements (with the indexes $j,j'=1,2,...,N$) given by  \cite{Mayuquan2010,Rezakhani2010}
\begin{equation}
Q_{\mu \nu}^{jj'}:=\langle\partial_{\lambda_{\mu}} \psi_{j}(\boldsymbol\lambda)|[1-P(\boldsymbol\lambda)]| \partial_{\lambda_{\nu}} \psi_{j'}(\boldsymbol\lambda)\rangle,
\end{equation}
where $P(\boldsymbol\lambda)=\sum_{j=1}^{N}\left|\psi_{j}(\boldsymbol\lambda)\right\rangle\left\langle\psi_{j}(\boldsymbol\lambda)\right|$ is the projection operator. The corresponding non-Abelian quantum metric $g_{\mu \nu}$ and Berry curvature $F_{\mu \nu}$ are given by
\begin{equation}
\begin{aligned}
g_{\mu \nu}&=(Q_{\mu \nu}+Q_{\mu \nu}^{\dagger})/2,\\
F_{\mu \nu}&=i(Q_{\mu \nu}-Q_{\mu \nu}^{\dagger}).
\end{aligned}
\end{equation}
Here $g_{\mu \nu}$ and $F_{\mu \nu}$ are $N\times N$ complex matrices for each $\mu$ and $\nu$, and satisfy the symmetric and antisymmetric relations $g_{\mu \nu}=g_{\mu \nu}^{\dagger}=g_{\nu \mu}$ and $F_{\mu \nu}=F_{\mu \nu}^{\dagger}=-F_{\nu \mu}$, respectively. Thus, the quantum metric tensor $g$ in the $U(N)$ non-Abelian case becomes the $DN\times DN$ matrix:
\begin{equation}\label{gMatrix}
g=\left(\begin{array}{lll}
g_{11}^{jj'} &  \cdots  & g_{1D}^{jj'}\\
~~\vdots &  \ddots  & ~~\vdots   \\
g_{D1}^{jj'} &  \cdots  & g_{DD}^{jj'}
\end{array}\right)_{DN\times DN}.
\end{equation}
The distance in Eq.~\eqref{nA-distance} can then be rewritten as
\begin{equation}
\label{SS}
\begin{aligned}
d S^{2} &=\sum_{\mu,\nu=1}^{D}\left[\left(c_{1}^{*},~\cdots,~c_{N}^{*}\right) g_{\mu \nu}\left(\begin{array}{c}
c_{1} \\
\vdots \\
c_{N}
\end{array}\right)\right] d \lambda_{\mu} d \lambda_{\nu}\\
&=\sum_{\mu,\nu=1}^{D} \left(\sum_{j,j'=1}^{N} c_{j}^{*}g_{\mu \nu}^{jj'}c_{j'} \right)d \lambda_{\mu} d \lambda_{\nu}.
\end{aligned}
\end{equation}
Note that the formulas in Eqs. (\ref{nA-distance}-\ref{SS}) naturally recover to those for the non-degenerate case when $N=1$. Even in the degenerate case, the distance $d S^{2}$ also corresponds to the transition probability from the degenerate ground state $|\Psi_{0}(\boldsymbol\lambda)\rangle$ to excited states after a sudden quench of the parameter from $\boldsymbol\lambda$ to $\boldsymbol\lambda+d\boldsymbol\lambda$. Thus, the sudden quench method \cite{TanXS2019,TanXS2021} can be generalized to extract the values of all the components of the non-Abelian quantum metric $g_{\mu \nu}^{jj'}$ and the related topological invariants.

\section{Extracting non-Abelian quantum metric from sudden quench}\label{sec3}
We proceed to show the non-Abelian quantum metric can be extracted by measuring the transition probability after sudden quenches of the Hamiltonian parameters. For the system Hamiltonian $H(\boldsymbol\lambda)$ initially prepared at $\boldsymbol\lambda=(\lambda_1,\lambda_2,...,\lambda_D)$, the components of the non-Abelian quantum metric tensor in Eq. (\ref{gMatrix}) at this point can be divided into four classes: the real diagonal parts $g_{\mu \mu}^{jj}$, the real off-diagonal parts $g_{\mu \nu}^{jj}$, and two complex off-diagonal $g_{\mu \mu}^{jj'}$ and $g_{\mu \nu}^{jj'}$ with $\mu\neq\nu$ and $j\neq j'$. To extract $g_{\mu \mu}^{jj}$, one can prepare the initial ground state $|\psi_{j}(\boldsymbol\lambda)\rangle$ and then perform a quantum quench, where the parameter suddenly changes from $\boldsymbol\lambda$ to $\boldsymbol\lambda+\delta \lambda \boldsymbol{e}_{\mu}$ with a small spacing $\delta \lambda$ along the $\boldsymbol{e}_{\mu}$ direction. After the sudden quench, the transition probability to the excited states is
\begin{equation}
\label{P1}
\Gamma_{\mu \mu}^{jj}=\sum_{m=N+1}^{N+M}\left|\left\langle\psi_{j}(\boldsymbol\lambda) | \psi_m (\boldsymbol\lambda+\delta \lambda\boldsymbol{e}_{\mu})\right\rangle\right|^{2}.
\end{equation}
By substituting the equation $|\psi_m(\boldsymbol\lambda+\delta \lambda)\rangle=|\psi_m(\boldsymbol\lambda)\rangle+|\partial_{\lambda_{\mu}} \psi_m(\boldsymbol\lambda)\rangle \delta \lambda+\mathcal{O}(\delta \lambda^2)$ into Eq.~\eqref{P1}, one can obtain the transition probability
\begin{equation}
\begin{aligned}
\Gamma_{\mu \mu}^{jj}
&=\langle\partial_{\lambda_{\mu}} \psi_{j}(\boldsymbol\lambda) | [1-P(\boldsymbol\lambda)] | \partial_{\lambda_{\mu}} \psi_{j}(\boldsymbol\lambda)\rangle  \delta\lambda^{2}+\mathcal{O}(\delta \lambda^3)\\
&=g_{\mu \mu}^{jj}\delta \lambda^2+\mathcal{O}(\delta \lambda^3).
\end{aligned}
\end{equation}
Thus, by measuring the transition probability $\Gamma_{\mu \mu}^{jj}$ from the initially state $|\psi_{j}(\boldsymbol\lambda)\rangle$ with different $j$ and along different quench directions $\boldsymbol{e}_{\mu}$, one can obtain all the diagonal components of the non-Abelian quantum metric at the point $\boldsymbol\lambda$ as
\begin{equation}\label{G1}
g_{\mu \mu}^{jj}\approx\Gamma_{\mu \mu}^{jj}/\delta \lambda^2.
\end{equation}

We then consider to extract the real components $g^{jj}_{\mu \nu}$ with $\mu\neq\nu$. To this end, the same initial ground state $|\psi_{j}(\boldsymbol\lambda)\rangle$ is prepared while the parameter is quenched from $\boldsymbol\lambda$ to $\boldsymbol\lambda+\delta \lambda \boldsymbol{e}_{\mu}+\delta \lambda \boldsymbol{e}_{\nu}$ along the $\boldsymbol{e}_{\mu}+\boldsymbol{e}_{\nu}$ direction. After the sudden quench, one can measure the resulting transition probability to excited states $\Gamma_{\mu \nu}^{jj}$, which is obtained as
\begin{equation}\label{P2}
\begin{aligned}
\Gamma_{\mu \nu}^{jj} &=\sum_{m=N+1}^{N+M}|\langle\psi_{j}(\boldsymbol\lambda) \mid \psi_m(\boldsymbol\lambda+\delta \lambda \boldsymbol{e}_{\mu}+\delta \lambda \boldsymbol{e}_{\nu})\rangle|^{2}\\
&=(g_{\mu \mu}^{jj}+g_{\nu \nu}^{jj}+2g_{\mu \nu}^{jj})\delta \lambda^2 + \mathcal{O}(\delta \lambda^3).
\end{aligned}
\end{equation}
Since the transition probabilities $\Gamma_{\mu \mu}^{jj}$ and $\Gamma_{\nu \nu}^{jj}$ have been previously measured to obtain the diagonal components $g_{\mu \mu}^{jj}$ and $g_{\nu \nu}^{jj}$, the off-diagonal components $g_{\mu \nu}^{jj}$ can then be extracted as
\begin{equation}\label{G2}
g_{\mu \nu}^{jj}\approx(\Gamma_{\mu \nu}^{jj}-\Gamma_{\mu \mu}^{jj}-\Gamma_{\nu \nu}^{jj})/(2\delta \lambda^2).
\end{equation}

To extract the third kind of components $g_{\mu \mu}^{jj'}$ ($j\neq j'$) that are generally complex \cite{Mayuquan2010}, we need to perform two independent quench protocols. One can first prepare the initial state $|\psi_{a}(\boldsymbol\lambda)\rangle=(|\psi_{j}(\boldsymbol\lambda)\rangle+|\psi_{j'}(\boldsymbol\lambda)\rangle)/ \sqrt{2}$ as a superposition of two degenerate ground eigenstates. After a sudden quench from $\boldsymbol\lambda$ to $\boldsymbol\lambda+\delta \lambda \boldsymbol{e}_{\mu}$, the transition probability to excited states is obtained as
\begin{equation}
\label{P33}
\Gamma_{\mu \mu}^{aa}=g_{\mu \mu}^{aa}\delta \lambda^2 + \mathcal{O}(\delta \lambda^3),
\end{equation}
where $g_{\mu \mu}^{aa}$ is the corresponding quantum metric component for the state $|\psi_{a}(\boldsymbol\lambda)\rangle$. To extract $g_{\mu \mu}^{jj'}$ for $|\psi_{j}(\boldsymbol\lambda)\rangle$, one can perform the same quench for another superposition initial state $|\psi_{b}(\boldsymbol\lambda)\rangle=(|\psi_{j}(\boldsymbol\lambda)\rangle+i|\psi_{j'}(\boldsymbol\lambda)\rangle)/ \sqrt{2}$. The resulting transition probability is given by
\begin{equation}
\label{P44}
\mathrm{\Gamma}_{\mu \mu}^{bb}=g_{\mu \mu}^{bb}\delta \lambda^2+ \mathcal{O}(\delta \lambda^3).
\end{equation}
From the derivations of $g_{\mu \mu}^{aa}$ and $g_{\mu \mu}^{bb}$, we can obtain
\begin{equation}
g_{\mu \mu}^{jj'}=\frac{2 i g_{\mu \mu}^{aa}+2 g_{\mu \mu}^{bb}-(1+i)\left(g_{\mu \mu}^{jj}+g_{\mu \mu}^{j'j'}\right)}{2 i}.
\end{equation}
The real and imaginary parts of the complex components $g_{\mu \mu}^{jj'}$ can finally be extracted from the measured transition probabilities $\Gamma_{\mu \mu}^{aa}$, $\Gamma_{\mu \mu}^{bb}$, $\Gamma_{\mu \mu}^{jj}$ and $\Gamma_{\mu \mu}^{j'j'}$:
\begin{equation}\label{G3}
\begin{aligned}
\text{Re}[g_{\mu \mu}^{jj'}]\approx\frac{2\Gamma_{\mu \mu}^{aa}-\Gamma_{\mu \mu}^{jj}-\Gamma_{\mu \mu}^{j'j'}}{2\delta \lambda^2},\\
\text{Im}[g_{\mu \mu}^{jj'}]\approx\frac{\Gamma_{\mu \mu}^{jj}+\Gamma_{\mu \mu}^{j'j'}-2\Gamma_{\mu \mu}^{bb}}{2\delta \lambda^2}.
\end{aligned}
\end{equation}
Similar quench protocols can be used to extract the last complex components $g_{\mu \nu}^{jj'}$ ($\mu\neq\nu$ and $j\neq j'$) by quenching the parameter from $\boldsymbol\lambda$ to $\boldsymbol\lambda+\delta \lambda \boldsymbol{e}_{\mu}+\delta \lambda \boldsymbol{e}_{\nu}$ along the $\boldsymbol{e}_{\mu}+\boldsymbol{e}_{\nu}$ direction. One can derive the following relation
\begin{equation}
g_{\mu \nu}^{jj'}=\frac{2 i g_{\mu \nu}^{aa}+2 g_{\mu \nu}^{bb}-(1+i)\left(g_{\mu \nu}^{jj}+g_{\mu \nu}^{j'j'}\right)}{2 i}.
\end{equation}
Thus, the real and imaginary parts of $g_{\mu \nu}^{jj'}$ can finally be extracted by measuring the transition probabilities $\Gamma_{\mu \nu}^{aa}$, $\Gamma_{\mu \nu}^{bb}$, $\Gamma_{\mu \nu}^{jj}$ and $\Gamma_{\mu \nu}^{j'j'}$:
\begin{equation}\label{G4}
\begin{aligned}
\text{Re}[g_{\mu \nu}^{jj'}]\approx\frac{2\Gamma_{\mu \nu}^{aa}-\Gamma_{\mu \nu}^{jj}-\Gamma_{\mu \nu}^{j'j'}}{2\delta \lambda^2},\\
\text{Im}[g_{\mu \nu}^{jj'}]\approx\frac{\Gamma_{\mu \nu}^{jj}+\Gamma_{\mu \nu}^{j'j'}-2\Gamma_{\mu \nu}^{bb}}{2\delta \lambda^2}.
\end{aligned}
\end{equation}
Up to now, we have shown how to extract all the components of the non-Abelian quantum metric tensor by using the sudden quench scheme. Note that this method is general and can be applied to arbitrary degenerate quantum states in parameter or momentum space of any dimensions. Thus, it could be implemented in engineered artificial quantum systems with tunable parameter spaces or Bloch bands \cite{TanXS2019,TanXS2021,MYu2020,MChen,Gianfrate2020,Asteria2019}.

\section{Applications}\label{sec4}
In this section, we present two prototypical applications to extract the related topological Chern numbers by measuring non-Abelian quantum metric in degenerate quantum systems. We demonstrate the quench protocols with numerical simulations in these two applications.

\subsection{Extracting the real Chern number}
We show that the real Chern number of a generalized Dirac point in $\mathcal{PT}$-symmetric Dirac semimetals \cite{ZhaoYX2017,WangK2020} can be extracted from the quantum-metric measurement. The Bloch Hamiltonian in three-dimensional momentum space of the real Dirac semimetals is given by
\begin{equation}
\label{H1}
\mathcal{H}_{3D}(\boldsymbol{k})=\hbar \Omega_0(d_x \alpha_x+d_y \alpha_y+d_z \alpha_z),
\end{equation}
where $\hbar\Omega_0$ is the irrelevant energy unit. The time unit is given by $2\pi/\Omega_0$, and we set $\hbar=\Omega_0=1$ in the following. $\boldsymbol{k}=(k_x,k_y,k_z)$ is the momentum, $d_{ x}=\sin k_{x}$, $d_{ y} =\sin k_{y}$, $d_{ z} =m_z-\cos k_{x}-\cos k_{y}-\cos k_{z}$ with $m_z$ as the dimensionless Zeeman strength, and $\alpha_{x}=\sigma_{3} \otimes \tau_{1}, \alpha_{y}=-\sigma_{1} \otimes \tau_{1}, \alpha_{z}=-\sigma_{0} \otimes \tau_{3}$ are $4\times4$ Dirac matrices. This Hamiltonian preserves both inversion ($\mathcal{P}$) symmetry and time reversal ($\mathcal{T}$) symmetry and certainly combined $\mathcal{PT}$ symmetry: $\mathcal{P}\mathcal{H}_{3D}(\boldsymbol{k})\mathcal{P}^{-1} =\mathcal{H}_{3D}(-\boldsymbol{k})$, $\mathcal{T}\mathcal{H}_{3D}(\boldsymbol{k})\mathcal{T}^{-1}=\mathcal{H}_{3D}(-\boldsymbol{k})$, and $\mathcal{PT}\mathcal{H}_{3D}(\boldsymbol{k})(\mathcal{PT})^{-1}=\mathcal{H}_{3D}(\boldsymbol{k})$. Here the corresponding symmetry operators $\mathcal{P}=\alpha_3$, $\mathcal{T}=\alpha_3 \mathcal{K}$, and $\mathcal{PT}=\mathcal{K}$ with $\mathcal{K}$ the complex conjugate operator. The $\mathcal{PT}$ symmetry implies that the Hamiltonian is real. For $m_z\in(0,3)$, the Hamiltonian describes topological semimetals with a pair of real Dirac points, acting as $Z_2$-type Dirac monopoles \cite{ZhaoYX2017}. We consider the case of $m_z=2$, where the two Dirac points located at $\boldsymbol{K}_{\pm}=(0,0,\pm \pi/2)$. Near the Dirac points $\boldsymbol{K}_{\pm}$, the low-energy effective Hamiltonians are
\begin{equation}
\label{Hn}
\mathcal{H}_{\text{eff},\pm}=q_x \alpha_x+q_y \alpha_y\pm q_z \alpha_z,
\end{equation}
$\textbf{q}_{\pm}=\textbf{k}-\textbf{K}_{\pm}$. Without loss of generality, we focus on the real Dirac point at $\boldsymbol{K}_{+}$ described by the effective Hamiltonian $\mathcal{H}_{\text{eff},+}=q_x \alpha_x+q_y \alpha_y+ q_z \alpha_z$. The corresponding energy spectrum is $E(\boldsymbol{q})=\pm \sqrt{q_x^2+q_y^2+q_z^2}$, where the two lower bands and two upper bands are degenerate, respectively. We can parameterize the momentum space as $q_{x}=q \sin \theta \cos \phi$, $q_{y}=q \sin \theta \sin \phi$, and $q_{z}=q \cos \theta$, where $q=|\boldsymbol{q}|$, $\theta\in(0,\pi]$ and $\phi\in(0,2\pi]$ are two spheral angles of a $\mathcal{S}^2$ sphere. For the two-fold degenerate ground states denoted by $|\psi_{1,2}(\boldsymbol{\lambda})\rangle$ with $\boldsymbol{\lambda}=(\theta,\phi)$ on the unit $\mathcal{S}^2$ sphere, the topological charge of the real Dirac monopole is characterized by the real Chern number \cite{ZhaoYX2017}
\begin{equation}
\label{v}
\mathcal{C}_{R}=\frac{1}{4 \pi} \int_{\mathcal{S}^2} d\theta d\phi \operatorname{tr}\left(I \mathcal{F}_{\theta\phi}^{R}\right)  \quad \bmod 2,
\end{equation}
where $I=-i\sigma_2$ is the generator of the $SO(2)$ group and $\mathcal{F}_{\theta\phi}^{R}=-iF_{\theta\phi}$ is the real Berry curvature. Here the non-Abelian Berry curvature defined for $|\psi_{1,2}(\theta,\phi)\rangle$ can be obtained as
\begin{equation}
F=\left(\begin{array}{ll}
F_{\theta\theta} & F_{\theta\phi}
\vspace{1ex}\\
F_{\phi\theta} & F_{\phi\phi}
\end{array}\right)=\frac{i\sin \theta}{2}\left(\begin{array}{cccc}
0 & 0 & 0 & 1
\vspace{1ex}\\
0 & 0 & -1  & 0
\vspace{1ex}\\
0 &  -1  & 0 & 0
\vspace{1ex}\\
1 & 0 & 0 & 0
\end{array}\right).
\end{equation}
The real Chern number for the real Dirac point $\boldsymbol{K}_{+}$ ($\boldsymbol{K}_{-}$) is obtained as $\mathcal{C}_{R}=1$ ($\mathcal{C}_{R}=-1$). The non-Abelian quantum metric is given by
\begin{equation} \label{g-realDirac}
g=\left(\begin{array}{ll}
g_{\theta\theta} & g_{\theta\phi}
\vspace{1ex}\\
g_{\phi\theta} & g_{\phi\phi}
\end{array}\right)=\left(\begin{array}{cccc}
\frac{1}{4} & 0 & 0 & 0
\vspace{1ex}\\
0 & \frac{1}{4} & 0 & 0
\vspace{1ex}\\
0 &0 & \frac{1}{4} \sin^2 \theta & 0
\vspace{1ex}\\
0 & 0 & 0 & \frac{1}{4} \sin ^{2} \theta
\end{array}\right).
\end{equation}
In this $U(2)$ non-Abelian case, we find the relation between the quantum metric and the Berry curvature as (see the Appendix A for the derivation.)
\begin{equation}
\label{F}
\begin{aligned}
F_{\theta \phi}^{12} =2 i \sqrt{\operatorname{det} g^{11}_{(\theta,\phi)}}=2 i \sqrt{\operatorname{det}\left(
                                                                                          \begin{array}{cc}
                                                                                            g^{11}_{\theta\theta} & g^{11}_{\theta\phi} \\
                                                                                            g^{11}_{\phi\theta} & g^{11}_{\phi\phi} \\
                                                                                          \end{array}                                                                                        \right)},\\
                                                                                          \vspace{1ex}
F_{\theta \phi}^{21} =-2 i \sqrt{\operatorname{det} g^{22}_{(\theta,\phi)}}=-2 i \sqrt{\operatorname{det}\left(
                                                                                          \begin{array}{cc}
                                                                                            g^{22}_{\theta\theta} & g^{22}_{\theta\phi} \\
                                                                                            g^{22}_{\phi\theta} & g^{22}_{\phi\phi} \\
                                                                                          \end{array}                                                                                        \right)}.
\end{aligned}
\end{equation}
The real Chern number can then be written as
\begin{equation}
\label{vg}
\begin{aligned}
\mathcal{C}_{R}=\frac{1}{2\pi} \int_{\mathcal{S}^2} d \theta d \phi\left(\sqrt{\operatorname{det} g^{11}_{(\theta,\phi)}}+\sqrt{\operatorname{det} g^{22}_{(\theta,\phi)}}\right) ~~\bmod 2.
\end{aligned}
\end{equation}
Thus, we can obtain $\mathcal{C}_{R}$ by measuring the non-Abelian quantum metric tensor $g$.

\begin{figure}
\centering
 \includegraphics[width=0.48\textwidth]{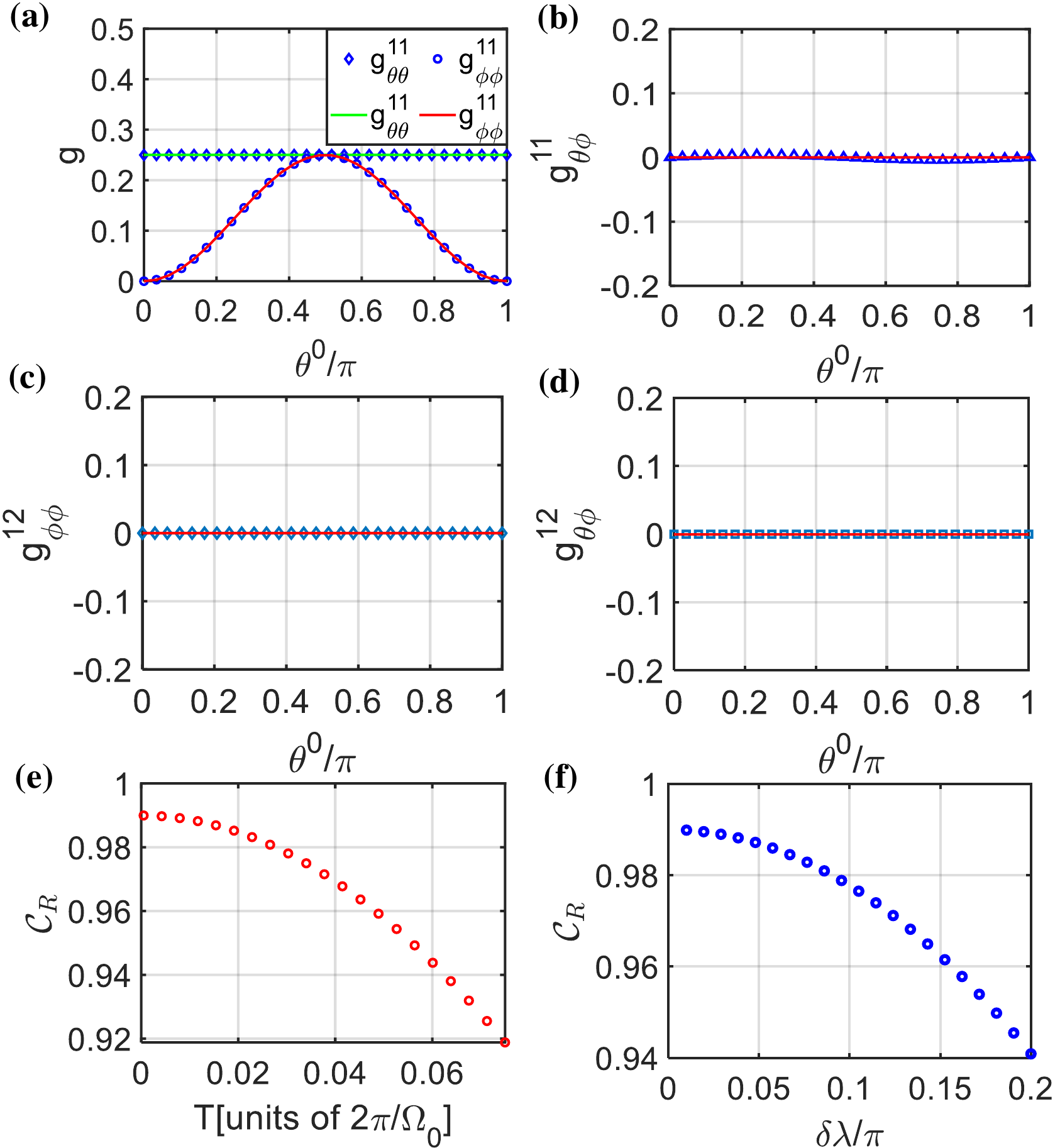}
 \caption{(Color online) Extraction of the non-Abelian quantum metric tensor in Eq.~\eqref{g-realDirac}. (a) $g_{\theta \theta}^{11}$, $g_{\phi \phi}^{11}$; (b) $g_{\theta \phi}^{11}$; (c) $g_{\phi \phi}^{12}$; and (d) $g_{\theta \phi}^{12}$ as a function of $\theta$ with fixed $\phi=\pi/4$. Dots and lines represent the numerical and theoretical results, respectively. The numerically obtained real Chern number $\mathcal{C}_R$ for different $T$ with fixed $\delta\theta=\delta\phi=\delta\lambda=\pi/100$ (e) and different $\delta\lambda$ with fixed $T=0.001$ (f). The numerical results are obtained from full-time-dynamics simulations.}
\label{fig1}
\end{figure}

We use the quench scheme to extract the components of $g$ in Eq. (\ref{g-realDirac}). We prepare the initial state at  $|\psi_{1}(\theta, \phi)\rangle$ and quench the parameter $\theta$ ($\phi$) to $\theta+\delta \theta$ along the $\boldsymbol{e}_{\theta}$ direction ($\phi+\delta\phi$ along the $\boldsymbol{e}_{\phi}$ direction) with fixed $\phi$ ($\theta$). The transition probability $\Gamma_{\theta \theta}^{11}$ ($\Gamma_{\phi \phi}^{11}$) to exciting states $|\psi_{3}\rangle$ and $|\psi_{4}\rangle$ is then measured, which gives the component $g^{11}_{\theta \theta}\approx\Gamma_{\theta \theta}^{11}/\delta \theta^2$ ($g^{11}_{\phi \phi}\approx\Gamma_{\phi \phi}^{11}/\delta \phi^2$). The other two diagonal components $g^{22}_{\theta \theta}$ and $g^{22}_{\phi \phi}$ can be obtained with the same protocol for the initial state $|\psi_{2}(\theta, \phi)\rangle$. To extract the off-diagonal components, such as $g_{\theta \phi}^{11}$, we can additionally measure the transition probability for the initial state $|\psi_{1}(\theta, \phi)\rangle$ with the parameters quenched from $(\theta, \phi)$ to $(\theta+\delta \theta,\phi+\delta \phi)$ along the $\boldsymbol{e}_{\theta}+\boldsymbol{e}_{\phi}$ direction. Other off-diagonal components, such as $g_{\phi \phi}^{12}$ and $g_{\theta \phi}^{12}$, can be further extracted from the superposition states $(|\psi_{1}\rangle+|\psi_{2}\rangle)/\sqrt{2}$ and $(|\psi_{1}\rangle+i|\psi_{2}\rangle)/\sqrt{2}$ by using Eq. (\ref{G3}) and Eq. (\ref{G4}). We use the full-time-dynamics simulations to demonstrate the quench scheme, with the numerical results of the extracted non-Abelian quantum metric shown in Fig.~\ref{fig1} (a-d), which agree well with the analytical results. In the simulations, we use the linear quench with $\boldsymbol\lambda(t)=\boldsymbol\lambda_0+t/T\delta\lambda\boldsymbol e_{\eta}$ along the $\boldsymbol e_{\eta}$ direction from the initial position $\boldsymbol\lambda_0$, where the quench time $T=0.001$ (in unit of $2\pi/\Omega_0$), $\delta \lambda=\delta\theta=\delta\phi=\pi/100$, and $\boldsymbol e_{\eta}=\{\boldsymbol e_{\theta}, \boldsymbol e_{\phi}, \boldsymbol e_{\theta}+\boldsymbol e_{\phi}\}$ for the corresponding protocols, respectively. With the numerically extracted non-Abelian quantum metric and Eq. (\ref{vg}), we obtain the real Chern number $\mathcal{C}_R=0.9899$, which is close to the theoretical value of $1$. In Figs.~\ref{fig1} (e) and (f), we show the numerical results of $\mathcal{C}_R$ for different $T$ and $\delta \lambda$, respectively. To ensure the validity of this quench scheme for extracting the real Chern number, sufficiently small values of $T$ and $\delta \lambda$ are required, which is similar to the case in Refs. \cite{TanXS2019,TanXS2021}.

\subsection{Extracting the second Chern number}

Now we show another application of non-Abelian quantum metric in measuring the second Chern number of a $SU(2)$ Yang monopole in five-dimensional space \cite{Yang}. To do this, we consider the generalized Dirac model Hamiltonian in five-dimension momentum space \cite{Abigail2018,Kolodrubetz2016}
\begin{equation}
\label{H-5D}
H_{5D}(\tilde{\boldsymbol{k}})=\hbar\Omega_0(\tilde{d}_1\beta_1+\tilde{d}_2\beta_2+\tilde{d}_3\beta_3+\tilde{d}_4\beta_4+\tilde{d}_5\beta_5),
\end{equation}
$\tilde{\boldsymbol{k}}=(k_x,k_y,k_z,k_w,k_v)$ is the momentum, $\tilde{d}_1=\sin k_{x}$, $\tilde{d}_2=\sin k_y$, $\tilde{d}_3=\sin k_z$, $\tilde{d}_4=\sin k_w$, $\tilde{d}_5=\tilde{m}_z-\cos k_x-\cos k_y-\cos k_z-\cos k_w-\cos k_v$, $\tilde{m}_z$ is the dimensionless Zeeman strength in this model. Here the $4\times4$ Dirac matrices are chosen as $\beta_1=\sigma_{0} \otimes \sigma_{3}$, $\beta_2=\sigma_{0} \otimes \sigma_{1}$, $\beta_3=-\sigma_{3} \otimes \sigma_{2}$, $\beta_4=\sigma_{2} \otimes \sigma_{2}$, and $\beta_5=\sigma_{1} \otimes \sigma_{2}$. For $\tilde{m}_z\in(0,5)$, the Hamiltonian describes a five-dimensional topological semimetal with a pair of Yang monopoles in $\tilde{\boldsymbol{k}}$ space \cite{XLQi2008,XLQi2011,BLian2016}. We consider the case of $\tilde{m}_z=4$, where the two Yang monopoles are located at $\tilde{\boldsymbol{K}}_{\pm}=(0,0,0,0,\pm \pi/2)$. Near the points $\tilde{\boldsymbol{K}_{\pm}}$, the low-energy effective Hamiltonians are
\begin{equation}
\label{H-Yang}
\tilde{\mathcal{H}}_{\text{eff},\pm}=\tilde{q}_x \beta_1+\tilde{q}_y \beta_2+ \tilde{q}_z \beta_3+\tilde{q}_w \beta_4 \pm \tilde{q}_v \beta_5,
\end{equation}
where $\tilde{\boldsymbol{q}}_{\pm}=\tilde{\boldsymbol{k}}-\tilde{\boldsymbol{K}}_{\pm}$. The time-reversal operator of the system is $T=\Theta K$, where $\Theta=i\sigma_{2} \otimes \sigma_0$ with $K$ being the complex conjugate operator. We consider the Hamiltonian $\tilde{\mathcal{H}}_{\text{eff},+}$ for the Yang monopole at $\tilde{\boldsymbol{K}}_{+}$. The corresponding four-band energy spectrum is $E(\tilde{\boldsymbol{k}})=\pm \sqrt{\tilde{q}_x^2+\tilde{q}_y^2+\tilde{q}_z^2+\tilde{q}_w^2+\tilde{q}_v^2}$, where the two lower (upper) bands are degenerate. The momenta can be parameterized as $\tilde{q}_{x}=\tilde{q} \cos \phi_1$, $\tilde{q}_{y}=\tilde{q} \sin \phi_1 \cos \phi_2$, $\tilde{q}_{z}=\tilde{q} \sin \phi_1 \sin\phi_2 \cos \phi_3$, $\tilde{q}_{w}=\tilde{q} \sin \phi_1 \sin\phi_2 \sin \phi_3 \cos \phi_4$, $\tilde{q}_{v}=\tilde{q} \sin \phi_1 \sin\phi_2 \sin \phi_3 \sin \phi_4$, where $\tilde{q}=|\tilde{\boldsymbol{q}}|$, $\phi_{1,2,3}\in(0,\pi]$, and $\phi_4\in(0,2\pi]$ are four spheral angles of an $\mathcal{S}^4$ sphere. For the two-fold degenerate ground states denoted by $|\psi_{1,2}(\tilde{\boldsymbol{\lambda}})\rangle$ with $\tilde{\boldsymbol{\lambda}}=(\phi_1,\phi_2,\phi_3,\phi_4)$ on the unit $\mathcal{S}^4$ sphere, the topological charge of the Yang monopole is characterized by the second Chern number \cite{XLQi2008,XLQi2011}
\begin{equation}
\begin{aligned}
\label{C2}
\mathcal{C}_{2}&=\frac{1}{32 \pi^{2}} \int_{\mathcal{M}^{4}} d^{4} x \epsilon^{\mu \nu \rho \sigma} \operatorname{tr}\left[F_{\mu \nu} F_{\rho \sigma}\right]\\
&=\frac{3}{4 \pi^{2}} \int_{\mathcal{S}^{4}} \operatorname{tr}\left[F_{\phi_{1} \phi_{2}} F_{\phi_{3} \phi_{4}}\right] d\phi_1 d\phi_2 d\phi_3 d\phi_4.
\end{aligned}
\end{equation}
Here the $U(2)$ non-Abelian Berry curvatures $F_{\phi_{1} \phi_{2}}$ and $F_{\phi_{3} \phi_{4}}$ can be obtained as
\begin{equation}
\begin{aligned}
&F_{\phi_{1} \phi_{2}}=\frac{i\sin\phi_1}{2}\left(\begin{array}{ll}
i \cos \phi_3 & - \sin\phi_3 \\
\sin \phi_3 & -i\cos\phi_3
\end{array}\right), \\
&F_{\phi_{3} \phi_{4}}=\frac{i\sin^2\phi_1\sin^2\phi_2}{4}\left(\begin{array}{ll}
-i\sin 2\phi_3& 2\sin^2\phi_3 \\
-2\sin^2\phi_3 & i\sin 2\phi_3
\end{array}\right).
\end{aligned}
\end{equation}
The topological charge of the Yang monopole at $\tilde{\boldsymbol{K}}_{+}$ can then be obtained as $\mathcal{C}_{2}=-1$. For the Yang monopole $\tilde{\boldsymbol{K}}_{-}$, one can obtain $\mathcal{C}_{2}=1$. Note that the first Chern numbers in this model are vanishing since $\operatorname{tr}[F_{\phi_{i} \phi_{j}}]=0~ (i,j=1,2,3,4)$ due to the time-reversal symmetry.

\begin{figure}
\centering
 \includegraphics[width=0.48\textwidth]{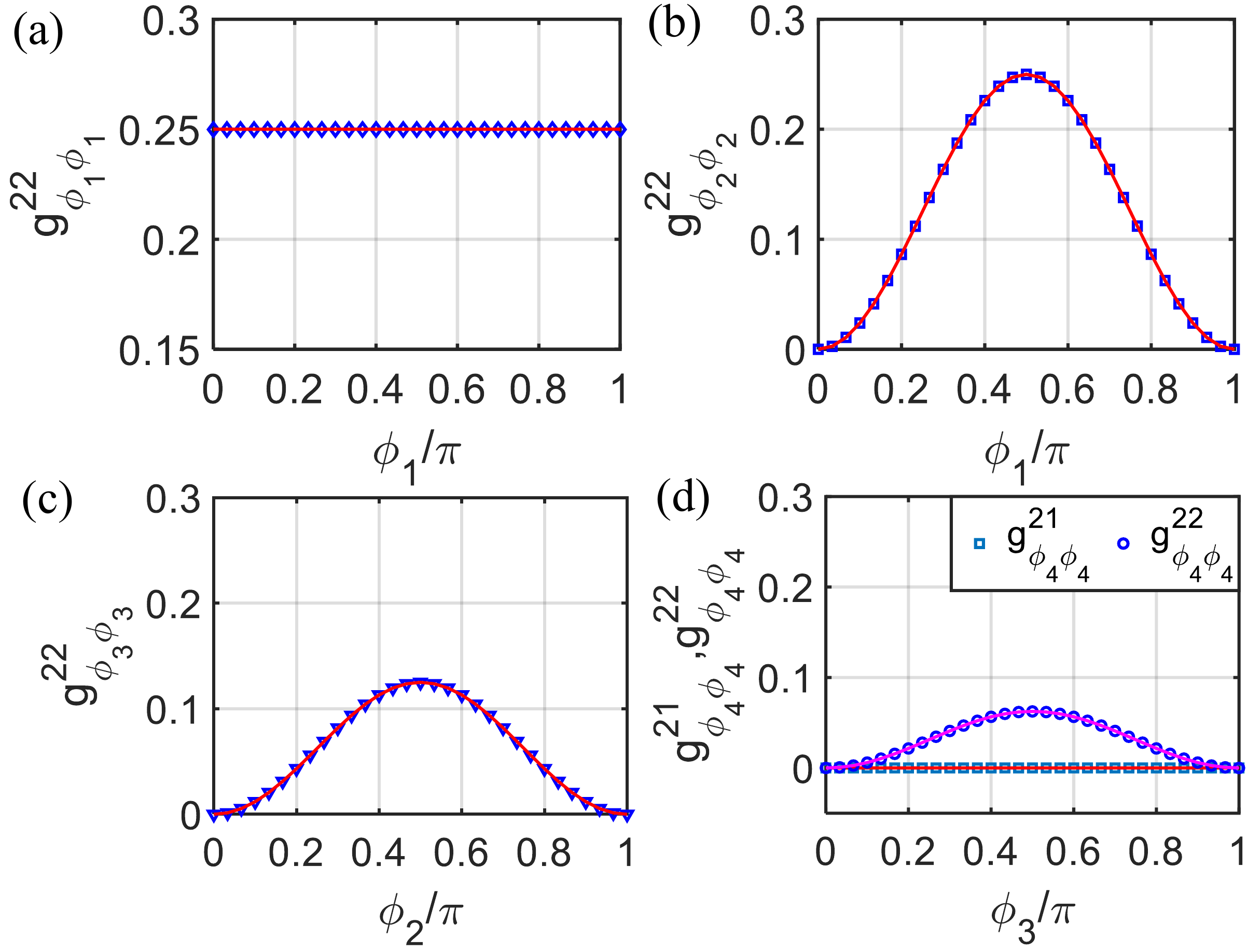}
 \caption{(Color online) Extraction of the non-Abelian quantum metric tensor in Eq.~\eqref{g-Yang}. (a) $g_{\phi_1\phi_1}^{22}$ as a function of $\phi_1$ for fixed $\phi_2=\phi_4=\pi/4, \phi_3=\pi$; (b) $g_{\phi_2 \phi_2}^{22}$ as a function of $\phi_1$ for fixed $\phi_2=\phi_4=\pi/4, \phi_3=\pi$; (c)$g_{\phi_3 \phi_3}^{22}$ as a function of $\phi_2$ for fixed $\phi_1=\phi_4=\pi/4, \phi_3=\pi$; (d) $g_{\phi_4 \phi_4}^{21}$ and $g_{\phi_4 \phi_4}^{22}$ as a function of $\phi_3$ for fixed $\phi_1=\phi_2=\phi_4=\pi/4$. Dots and lines represent the numerical and theoretical results, respectively. The numerical results are obtained from full-time-dynamics simulations.}
\label{fig2}
\end{figure}

In this model, the corresponding non-Abelian quantum metric tensor $g$ is an $8 \times 8$ matrix:
\begin{equation}
\label{g-Yang}
\begin{aligned}
&g=\left(\begin{array}{cccc}
g_{\phi_1 \phi_1} & g_{\phi_1 \phi_2} & g_{\phi_1 \phi_3} & g_{\phi_1 \phi_4} \\
g_{\phi_2 \phi_1} & g_{\phi_2 \phi_2} & g_{\phi_2 \phi_3} & g_{\phi_2 \phi_4} \\
g_{\phi_3 \phi_1} & g_{\phi_3 \phi_2} & g_{\phi_3 \phi_3} & g_{\phi_3 \phi_4} \\
g_{\phi_4 \phi_1} & g_{\phi_4 \phi_2} & g_{\phi_4 \phi_3} & g_{\phi_4 \phi_4}
\end{array}\right),   \\
&g_{\phi_i \phi_j}=\left(\begin{array}{ll}
g_{\phi_i \phi_j}^{11} & g_{\phi_i \phi_j}^{12} \\
g_{\phi_i \phi_j}^{21} & g_{\phi_i \phi_j}^{22}
\end{array}\right).
\end{aligned}
\end{equation}
We obtain the nonzero components $g_{\phi_1 \phi_1}^{11}=g_{\phi_1 \phi_1}^{22}=1/4$, $g_{\phi_2 \phi_2}^{11}=g_{\phi_2 \phi_2}^{22}=\sin^2 \phi_1/4$, $g_{\phi_3 \phi_3}^{11}=g_{\phi_3 \phi_3}^{22}=\sin^2 \phi_2 \sin^2 \phi_1/4$, $g_{\phi_4 \phi_4}^{11}=g_{\phi_4 \phi_4}^{22}=\sin^2 \phi_3\sin^2 \phi_2 \sin^2 \phi_1/4$, which are the diagonal components. All the  off-diagonal components are zero here.

It has been recently shown that the quantum metric is related to the Berry curvature in this model as \cite{AWZhang2021,BMera2021}
\begin{equation}
\sqrt{\operatorname{det}G}=\frac{1}{48}\left|\mathcal{F}\right|,
\end{equation}
where $G$ is a $4\times4$ matrix closely related to $g$ with the matrix elements $G_{ij}=\operatorname{tr}(g_{\phi_i\phi_j})$ with $i,j=\{1,2,3,4\}$ in Eq. (\ref{g-Yang}), and $\mathcal{F}=\epsilon^{a b c d} \operatorname{tr}\left(F_{a b} F_{c d}\right)$ with $a,b,c,d=\{\phi_1,\phi_2,\phi_3,\phi_4\}$. The second Chern number of the Yang monopole at $\tilde{\boldsymbol{K}}_{+}$ can then be given by \cite{AWZhang2021,BMera2021}
\begin{equation}
\begin{aligned}
\label{C2}
\mathcal{C}_{2}&=\frac{3}{\pi^{2}} \int_{\mathcal{S}^{4}} \operatorname{sgn}(\mathcal{F}) \sqrt{\operatorname{det}G} ~d\phi_1 d\phi_2 d\phi_3 d\phi_4\\
&=-\frac{3}{\pi^{2}} \int_{\mathcal{S}^{4}} \sqrt{\operatorname{det}G}~d\phi_1 d\phi_2 d\phi_3 d\phi_4.
\end{aligned}
\end{equation}
Thus, the second Chern number $\mathcal{C}_{2}$ can be extracted by measuring the non-Abelian quantum metric (and the sign of $\mathcal{F}$ for determining the sign of $\mathcal{C}_{2}$). Note that the non-Abelian Berry curvature has been experimentally measured to extract the second Chern number from the linear response to slowly varying the parameters of the system Hamiltonian \cite{Abigail2018}. Our scheme provides another way to extract the second Chern number from the quantum-metric measurements with short quench time, such that the decoherence effect in practical experiments would be negligible \cite{TanXS2019,TanXS2021}. Very recently, a time-periodic modulation method is proposed to extract second Chern number from measuring the sum of the non-Abelian quantum metric tensor $g_{a b} \equiv \sum_{n \in o c c} g_{a b}^{n n}$ \cite{AWZhang2021}, which can not be used to probe all components of the non-Abelian quantum metric, such as $g_{xy}^{nn}$.

\begin{figure}
\centering
 \includegraphics[width=0.48\textwidth]{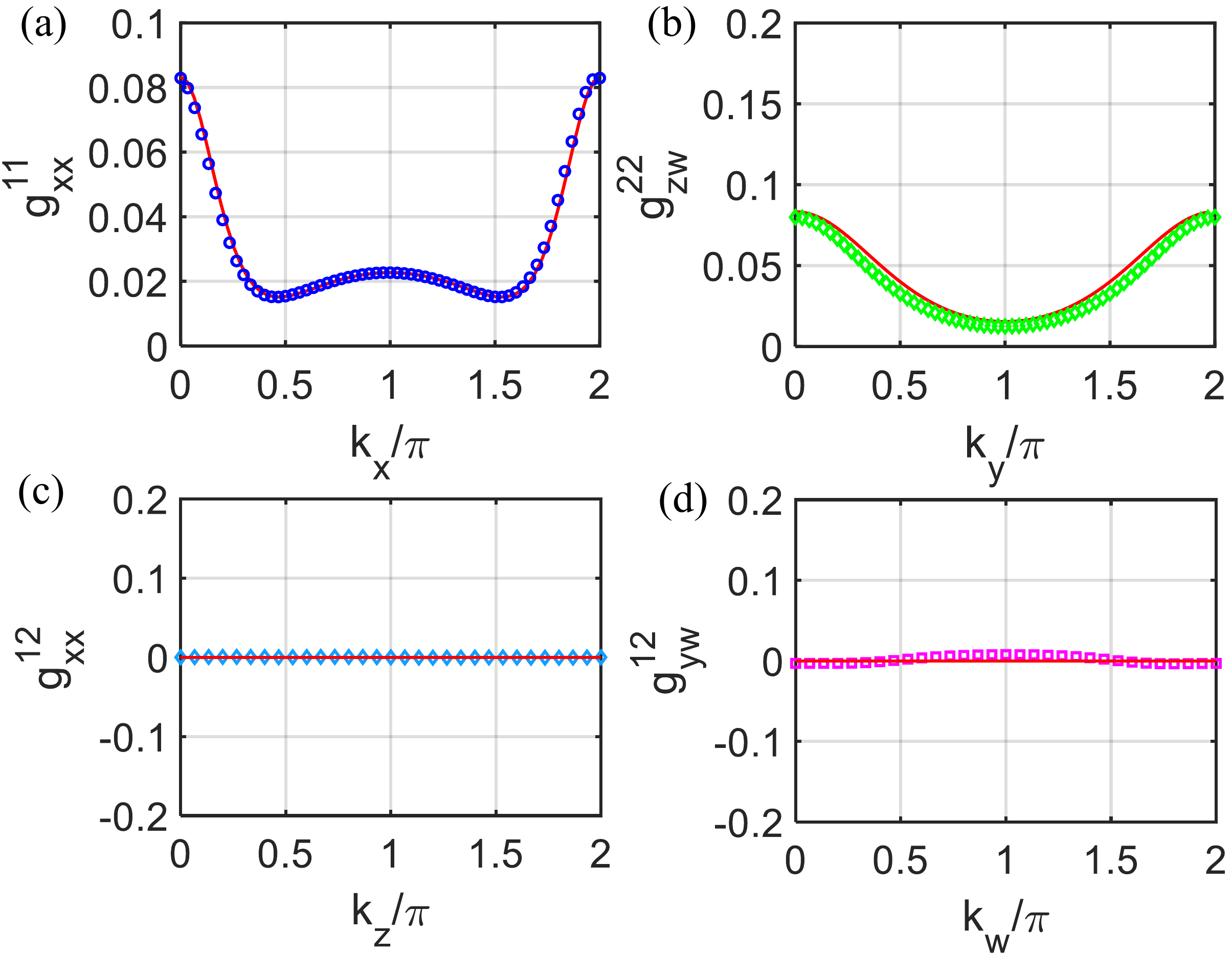}
\caption{(Color online) Extraction of the non-Abelian quantum metric tensor. (a) $g_{xx}^{11}$ as a function of $k_x$ for fixed $k_y=k_z=\pi/2, k_w=\pi$; (b) $g_{zw}^{22}$ as a function of $k_y$ for fixed $k_x=k_z=k_w=\pi/2$; (c) $g_{xx}^{12}$ as a function of $k_z$ for fixed $k_x=0, k_y=k_w=\pi/2$; (d) $g_{yw}^{12}$ as a function of $k_w$ for fixed $k_x=k_y=\pi/2, k_z=\pi/4$. Dots and lines represent the numerical and theoretical results, respectively. The numerical results are obtained from full-time-dynamics simulations.}
\label{fig3}
\end{figure}

All the components of non-Abelian quantum metric $g$ in Eq. (\ref{g-Yang}) can be extracted by using the quench scheme. For the diagonal components $g_{\phi_j \phi_j}^{11}$ and $g_{\phi_j \phi_j}^{22}$, we prepare the initial state at $|\psi_{1}\rangle$ and $|\psi_{2}\rangle$, respectively. Then we quench the parameter $\phi_j$ to $\phi_j+\delta \phi_j$ along $\boldsymbol{e}_{\phi_j}$ direction. The corresponding transition probabilities $\Gamma_{\phi_j \phi_j}^{11}$ and $\Gamma_{\phi_j \phi_j}^{22}$ to $|\psi_{3}\rangle$ and $|\psi_{4}\rangle$ can be measured to obtain $g_{\phi_j \phi_j}^{11}\approx \Gamma_{\phi_j \phi_j}^{11}/\delta\phi_j^2$ and $g_{\phi_j \phi_j}^{22}\approx \Gamma_{\phi_j \phi_j}^{22}/\delta\phi_j^2$. To extract the off-diagonal components $g_{\phi_i \phi_j}^{11}$, we should measure the transition probability $\Gamma_{\phi_i \phi_j}^{11}$ for the initial state $|\psi_{1}\rangle$ with the parameters quenched form $(\phi_i,\phi_j)$ to $(\phi_i+\delta \phi_i,\phi_j+\delta \phi_j)$ along the $\boldsymbol{e}_{\phi_i}+\boldsymbol{e}_{\phi_j}$ direction. The components $g_{\phi_i \phi_j}^{22}$ are obtained with the same protocol for the initial state $|\psi_{2}\rangle$. Other off-diagonal components $g_{\phi_j \phi_j}^{12}$, $g_{\phi_j \phi_j}^{21}$, $g_{\phi_i \phi_j}^{12}$ and $g_{\phi_i \phi_j}^{21}$ can be extracted from the superposition states $(|\psi_{1}\rangle+|\psi_{2}\rangle)/\sqrt{2}$ and $(|\psi_{1}\rangle+i|\psi_{2}\rangle)/\sqrt{2}$ by using Eq. (\ref{G3}) and Eq. (\ref{G4}). We perform the full-time-dynamics simulations to extract several typical components, with the numerical results shown in Fig.~\ref{fig2}, which agree well with the analytical results. In the simulations, we use the linear quench with $\tilde{\boldsymbol\lambda}(t)=\tilde{\boldsymbol\lambda}_0+t/T\delta\tilde{\lambda}\boldsymbol e_{\xi}$ along the $\boldsymbol e_{\xi}$ direction from the initial position $\tilde{\boldsymbol\lambda}_0$, where the quench time $T=0.001$ (in unit of $2\pi/\Omega_0$), $\delta \tilde{\lambda}=\delta\phi_1=\delta\phi_2=\delta\phi_3=\delta\phi_4=\pi/80$, and $\boldsymbol e_{\xi}=\{\boldsymbol e_{\phi_1}, \boldsymbol e_{\phi_2}, \boldsymbol e_{\phi_3}, \boldsymbol e_{\phi_4}\}$ for the corresponding protocols, respectively. With the numerically extracted non-Abelian quantum metric and Eq. (\ref{C2}), we obtain the second Chern number $\mathcal{C}_2=-0.9627$, which is close to the theoretical value of $-1$.

Finally, we consider the four-dimensional subsystem described by the Hamiltonian in Eq. (\ref{H-5D}) with $\tilde{m}_z=1$ and $k_v=\pi/2$, which corresponds to a topological insulator in $\tilde{\boldsymbol{k}}'=(k_x,k_y,k_z,k_w)$ space with $k_{x,y,z,w}\in(0,2\pi]$. The corresponding non-Abelian quantum metric are denoted by $g_{\mu\nu}^{jj'}$ with $\mu,\nu=x,y,z,w$ as the momentum indexes and $j,j'=1,2$ as the ground states indexes, which can also be extracted by using similar quench protocols. Fig.~\ref{fig3} shows the numerical results of several typical components of the quantum metric from the full-time-dynamics simulations. The numerical results agree well with the analytical ones, which indicates the feasibility of the quench scheme.

\begin{figure}
\centering
 \includegraphics[width=0.4\textwidth]{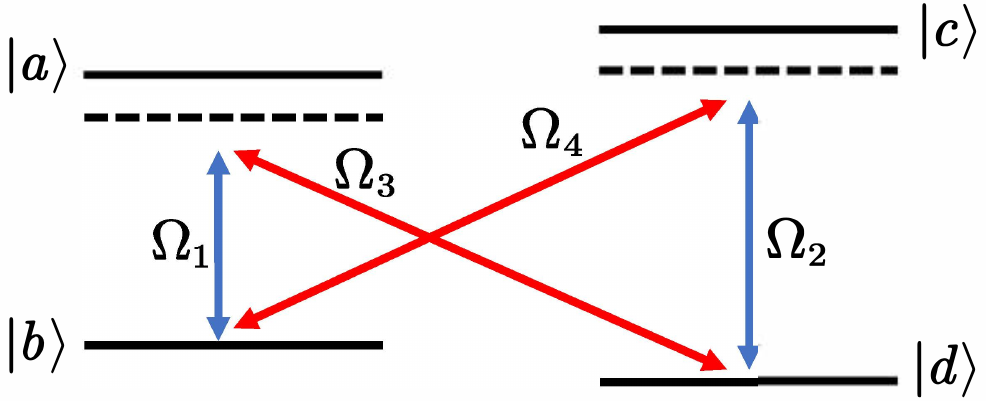}
\caption{(Color online) Diagrammatic sketch of a four-level atomic system for simulating the Hamiltonians with the form of Eqs. (\ref{H1}) and (\ref{Hn}) in the parameter space. }
\label{fig4}
\end{figure}

\section{Discussion and conclusion}\label{sec5}

Before concluding, we present a concrete experimental platform to simulate the Hamiltonians with the form of Eq. (\ref{H1}) and Eq. (\ref{Hn}) in the parameter space, following the manipulation of four-level ${ }^{87} \mathrm{Rb}$ atoms in Ref. \cite{QXL2021}. These Hamiltonians can be realized with a four-level atomic system as shown in Fig.~4. In a ${ }^{87} \mathrm{Rb}$ atomic system, we can choose the following four atomic levels: $\left|a\right\rangle=\left|F=2, m_{F}=-1\right\rangle$, $\left|b\right\rangle=\left|F=1, m_{F}=-1\right\rangle$, $\left|c\right\rangle=\left|F=2, m_{F}=0\right\rangle$ and $\left|d\right\rangle=\left|F=1, m_{F}=0\right\rangle$.  Using the bare state basis $\{ |a\rangle,|b\rangle,|c\rangle,|d\rangle    \}$, the Hamiltonian is given by
\begin{equation}
\begin{aligned}
H^{\prime}\!=&\!\left(\omega_{a}-\omega_{b}\right)\left|a\right\rangle\left\langle a\right|\!+\!\left(\omega_c-\omega_b\right)\left|c\right\rangle\left\langle c\right|\!+\!\left(\omega_d-\omega_b\right)\left|d\right\rangle\left\langle d\right| \\
&+\left(\Omega_{1}e^{i \omega_{1} t} e^{i \varphi_{1}}\left|a\right\rangle\left\langle b\right|\right.+\Omega_{2}e^{i \omega_{2} t} e^{i \varphi_{2}}\left|c\right\rangle\left\langle d\right| \\
&+\Omega_{3}e^{i \omega_{3} t} e^{i \varphi_{3}}\left|a\right\rangle\left\langle d\right| \left.+\Omega_{4}e^{i \omega_{4} t} e^{i \varphi_{4}}\left|c\right\rangle\left\langle b\right|+H.c.\right),
\end{aligned}
\end{equation}
where $\omega_{i} (i=a,b,c,d)$ is the energy frequencies of $\left|i\right\rangle$. $\Omega_{l}, \omega_{l}, \varphi_{l}$ correspond to the Rabi frequencies, frequencies and phases of the controlling microwaves, respectively. In the interaction picture, the Hamiltonian becomes $H_I=U H^{\prime} U^{\dagger}-i\left(\partial_{t} U^{\dagger}\right) U,$ where $U=e^{-i \omega_{1} t}\left|a\right\rangle\left\langle a|+|b\right\rangle\left\langle b\right|+e^{-i \omega_{4} t}|c\rangle\langle c|+e^{-i(\omega_{1}-\omega_{3}) t} |d\rangle\langle d|.$
Using the rotating-wave approximation, we can simplify the Hamiltonian as below $(\hbar=1)$
\begin{equation}
H_{\text{exp}}\!=\!\left(\!\!\begin{array}{cccc}
-\Delta_{1} & \Omega_{1}e^{-i \varphi_{1}} & 0 & \Omega_{3}e^{-i \varphi_{3}} \\
\Omega_{1}e^{i \varphi_{1}} & 0 & \Omega_{4}e^{i \varphi_{4}} & 0 \\
0 & \Omega_{4}e^{-i \varphi_{4}} & -\Delta_{4} & \Omega_{2}e^{-i \varphi_{2}} e^{-i \Delta^{\prime} t} \\
\Omega_{3}e^{i \varphi_{3}} & 0 & \Omega_{2}e^{i \varphi_{2}} e^{i \Delta^{\prime} t} & \Delta_{3}-\Delta_{1}
\end{array}\!\!\right)
\end{equation}
where
$\Delta^{\prime}=\omega_1+\omega_2-\omega_3-\omega_4=\Delta_1+\Delta_2-\Delta_3-\Delta_4$, $\Delta_1=\omega_1-(\omega_a-\omega_b)$, $\Delta_2=\omega_2-(\omega_c-\omega_d)$, $\Delta_3=\omega_3-(\omega_a-\omega_d)$, $\Delta_4=\omega_4-(\omega_c-\omega_b)$. The Hamiltonian in Eq. (\ref{H1}) can be derived  if we set $\Delta_1=\Delta_2=\Delta_3=\Delta_4=\Delta=\Omega_0 d_z$ and lift the energy levels by $\Delta$, $\varphi_1=\varphi_2=\varphi_3=\varphi_4=\Delta^{\prime}=0$, $\left\{\Omega_1,\Omega_2,\Omega_3,\Omega_4\right\}=\left\{\Omega_0d_x,-\Omega_0d_x,-\Omega_0d_y,-\Omega_0d_y\right\}$. On the other hand,  the Hamiltonian in Eq.~\eqref{Hn} can be constructed if the parameters become $\left\{\Omega_1,\Omega_2,\Omega_3,\Omega_4\right\}=\Omega_0\left\{\sin\theta\cos\phi,-\sin\theta\cos\phi,-\sin\theta\sin\phi,-\sin\theta\sin\phi\right\}$, $\Delta_1=\Delta_2=\Delta_3=\Delta_4=\Delta=\pm\Omega_0\cos\theta$.

In summary, we have proposed an experimental scheme to directly extract the non-Abelian quantum metric tensor of degenerate quantum states based on measuring the transition probabilities after parameter quenches. We have shown that the non-Abelian quantum metric can be used to obtain the real Chern number of an $O(2)$-generalized Dirac monopole and the second Chern number of an $SU(2)$ Yang monopole, which can be simulated in three and five-dimensional parameter space of artificial quantum systems, respectively. We have further demonstrated our quench scheme for these two applications with numerical simulations.

\acknowledgments
This work was supported by the Key-Area Research and Development Program of Guangdong Province (Grant No. 2019B030330001), the National Natural Science Foundation of China (Grants No. 12174126, No. 12074180, and No. U1801661), the Science and Technology Program of Guangzhou (Grants No. 2019050001), and the Guangdong Basic and Applied Basic Research Foundation (Grant No. 2021A1515010315).

{\sl Note added}. - After submission of this manuscript, we noticed a recent work on proposing to measure quantum geometric tensor in non-Abelian systems by using Rabi oscillations \cite{Weisbrich2021}.

\begin{appendix} \label{app}

\section{Derivation of Eq. (\ref{F})}
Four orthonormal eigenstates of the Hamiltonian $\mathcal{H}_{\text{3D}}$ in Eq.~\eqref{H1} with $E_{\pm}=\pm \sqrt{d_{x}^2+d_{y}^2+d_{z}^2}$ are given by
\begin{equation}
\begin{array}{l}
\left|\beta_{1}\right\rangle=N_{-}\left(\begin{array}{c}
-d_{x} \\
d-d_{z} \\
d_{y} \\
0
\end{array}\right), \quad\left|\beta_{2}\right\rangle=N_{-}\left(\begin{array}{c}
d_{y} \\
0 \\
d_{x} \\
d-d_{z}
\end{array}\right), \\
\left|\beta_{3}\right\rangle=N_{+}\left(\!\!\begin{array}{c}
-d_{x} \\
-d-d_{z} \\
d_{y}\\
0
\end{array}\!\!\right), \quad\left|\beta_{4}\right\rangle=N_{+}\left(\begin{array}{c}
-d_{y} \\
0 \\
-d_{x} \\
d+d_{z}
\end{array}\right).
\end{array}
\end{equation}
Here $N_{\pm}=1/ \sqrt{2d^2 \pm 2dd_z}$ and $d=E_{+}$.
Then it can be easily verified that
$Q_{\mu\mu}^{12}=0, Q_{\mu\nu}^{12}=-Q_{\nu\mu}^{12}, Q_{\mu\nu}^{11}=Q_{\nu\mu}^{11}=0, Q_{\mu\nu}^{22}=Q_{\nu\mu}^{22}=0, \text{with} (\mu,\nu)=(x,y)$. In this case, the quantum geometric tensor can be written as
\begin{equation}
Q=\left(\begin{array}{cccc}
g_{xx}^{11} & 0 & g_{xy}^{11} & -i \frac{F_{xy}^{12}}{2} \\
0 & g_{xx}^{22} & -i \frac{F_{xy}^{21}}{2} & g_{xy}^{22} \\
g_{yx}^{11} & -i \frac{F_{yx}^{12}}{2} & g_{yy}^{11} & 0 \\
-i \frac{F_{yx}^{21}}{2} & g_{yx}^{22} & 0 & g_{yy}^{22}
\end{array}\right).
\end{equation}
We also find that $Q_{yx}Q_{yy}=Q_{yy}Q_{yx}$, then the determinant of the non-Abelian quantum geometric tensor can be written as $\det Q=Q_{xx}Q_{yy}-Q_{xy}Q_{yx}$, i.e.,
\begin{equation}
\operatorname{det} Q=\left[\operatorname{det} g^{11}_{(x,y)}+\left(\frac{F_{xy}^{12}}{2}\right)^{2}\right]\!\!\!\left[\operatorname{det} g^{22}_{(x,y)}+\left(\frac{F_{xy}^{21}}{2}\right)^{2}\right],
\end{equation}
where
\begin{equation}
g^{11}_{(x,y)}=\left(\begin{array}{ll}
g^{11}_{xx} & g^{11}_{xy} \\
g^{11}_{yx} & g^{11}_{yy}
\end{array}\right), \quad g^{22}_{(x,y)}=\left(\begin{array}{ll}
g^{22}_{xx} & g^{22}_{xy} \\
g^{22}_{yx} & g^{22}_{yy}
\end{array}\right).
\end{equation}
For the $\mathcal{PT}$-symmetric Hamiltonian in Eq.~\eqref{H1}, $\det Q=0$ and the components of the non-Abelian Berry curvature $F_{xy}^{12}=i(Q_{xy}^{12}-Q_{yx}^{12})$ and $F_{xy}^{21}=i(Q_{xy}^{21}-Q_{yx}^{21})$ are imaginary numbers, then
\begin{equation}
\left|F_{x y}^{12}\right|=2 i \sqrt{\operatorname{det} g^{11}_{(x,y)}}, \quad\left|F_{x y}^{21}\right|=2 i \sqrt{\operatorname{det} g^{22}_{(x,y)}}.
\end{equation}
For the Hamiltonian $\mathcal{H}_{\text{eff}, +}$ in Eq.~\eqref{Hn}, there is a real Dirac monopole, the components of the non-Abelian curvature $F_{\theta\phi}^{12}>0$ and $F_{\theta\phi}^{21}<0$ for all the range of the parameters, then we can derive
\begin{equation}
\label{FF}
\begin{aligned}
F_{\theta \phi}^{12} =2 i \sqrt{\operatorname{det} g^{11}_{(\theta,\phi)}}=2 i \sqrt{\operatorname{det}\left(
                                                                                          \begin{array}{cc}
                                                                                            g^{11}_{\theta\theta} & g^{11}_{\theta\phi} \\
                                                                                            g^{11}_{\phi\theta} & g^{11}_{\phi\phi} \\
                                                                                          \end{array}                                                                                        \right)},\\
                                                                                          \vspace{1ex}
F_{\theta \phi}^{21} =-2 i \sqrt{\operatorname{det} g^{22}_{(\theta,\phi)}}=-2 i \sqrt{\operatorname{det}\left(
                                                                                          \begin{array}{cc}
                                                                                            g^{22}_{\theta\theta} & g^{22}_{\theta\phi} \\
                                                                                            g^{22}_{\phi\theta} & g^{22}_{\phi\phi} \\
                                                                                          \end{array}                                                                                        \right)}.
\end{aligned}
\end{equation}
Equation (B6) is just Eq.~\eqref{F}.
\end{appendix}

\end{document}